# A Characterization of Markov Equivalence Classes for Directed Acyclic Graphs with Latent Variables

Jiji Zhang
Division of Humanities and Social Sciences
California Institute of Technology
Pasadena, CA 91125
jiji@hss.caltech.edu

## Abstract

Different directed acyclic graphs (DAGs) may be Markov equivalent in the sense that they entail the same conditional independence relations among the observed variables. Meek (1995) characterizes Markov equivalence classes for DAGs (with no latent variables) by presenting a set of orientation rules that can correctly identify *all* arrow orientations shared by all DAGs in a Markov equivalence class, given a member of that class. For DAG models with latent variables, maximal ancestral graphs (MAGs) provide a neat representation that facilitates model search. Earlier work (Ali et al. 2005) has identified a set of orientation rules sufficient to construct all arrowheads common to a Markov equivalence class of MAGs. In this paper, we provide extra rules sufficient to construct all common tails as well. We end up with a set of orientation rules sound and complete for identifying commonalities across a Markov equivalence class of MAGs, which is particularly useful for causal inference.

## 1 INTRODUCTION

Directed acyclic graphs (DAGs) are now widely used both as statistical models and as causal models. Different DAGs represent different causal structures, but may be Markov equivalent in the sense that they entail the same conditional independence relations among the observed variables, and hence cannot be distinguished by observed patterns of independence and dependence. So it is important for the sake of causal inference to characterize common features of Markov equivalent causal models. Characterizations of Markov equivalence classes of DAGs (with no latent variables) are available in the literature (Meek 1995,

Chickering 1995, Andersson et al. 1997). In particular, Meek (1995) gave a set of orientation rules that are sound and complete for identifying arrow orientations common to all DAGs in a Markov equivalence class, given a member of that class.

In many cases, however, the data generating process might involve unobserved confounders or selection variables, and we need to consider DAGs with latent variables to model the process. Such latent variable DAG models can be represented by ancestral graphical models (Richardson and Spirtes - henceforth RS, 2002), in that for any DAG with latent variables, there is a unique maximal ancestral graph (MAG) that represents the conditional independence relations and causal relations among the observed variables entailed by the DAG. Ali et al. (2005) made an important step towards characterizing Markov equivalence classes for MAGs by providing rules sufficient to construct all arrowheads common to a Markov equivalence class of MAGs. In this paper, we provide extra rules sufficient to construct all common tails, which encode important causal information. We end up with a set of orientation rules sound and complete for identifying commonalities across a Markov equivalence class of MAGs.

Section 2 introduces relevant definitions. We summarize the arrowhead complete rules in section 3, and present the extra rules and tail completeness in section 4. We close with some discussions in section 5.

## 2 BACKGROUND

The following example attributed to Chris Meek in Richardson (1998) illustrates nicely the primary motivation behind ancestral graphs:

> "The graph [Figure 1] represents a randomized trial of an ineffective drug with unpleasant side-effects. Patients are randomly assigned to the treatment or control group ($A$). Those in the treatment group suffer unpleasant side-effects ($Ef$), the severity of which is influenced by the patient's general level of health ($H$), with sicker patients suffering worse side-effects. Those patients who suffer sufficiently severe side-effects are likely to drop out of the study. The selection variable ($Sel$) records whether or not a patient remains in the study, thus for all those remaining in the study $Sel = StayIn$. Since unhealthy patients who are taking the drug are more likely to drop out, those patients in the treatment group who remain in the study tend to be healthier than those in the control group. Finally health status ($H$) influences how rapidly the patient covers ($R$)." (Richardson 1998, pp. 234)

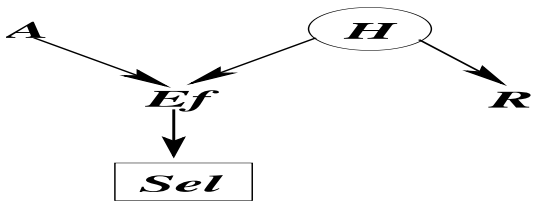

Figure 1: A Causal Mechanism with Latent and Selection Variables

This simple case shows how the presence of latent confounders and/or selection variables matters. The variables of primary interest, $A$ and $R$, are observed to be correlated, even though the supposed causal mechanism entails independence between them. This correlation is not due to sample variation, but rather corresponds to genuine probabilistic association induced by design – only the subjects that eventually stay in the study are considered. The observed correlation is in effect a correlation conditional on the selection variable $Sel$, a canonical example of *selection effect*. $H$, on the other hand, is a familiar latent confounder that contributes to "spurious correlation".

A major virtue of ancestral graphs is that, without explicitly including latent variables, they can represent conditional independence relations and causal relations among observed variables when the underlying data generating process involves latent confounders and/or selection variables (Spirtes and Richardson 1996). This of course requires a richer syntax than DAGs. Besides directed edges, an ancestral graph can also contain bi-directed edges (associated with the presence of latent confounders), and undirected edges (associated with the presence of selection variables).

## 2.1 MAXIMAL ANCESTRAL GRAPHS

By a **mixed graph** we denote a vertex-edge graph that can contain three kinds of edges: directed ($\rightarrow$), bi-directed ($\leftrightarrow$) and undirected ($-$). The two ends of an edge we call **marks** or **orientations**. So the two marks of a bi-directed edge are both **arrowheads** ($>$), the two marks of an undirected edge are both **tails** ($-$), and a directed edge has one of each. We say an edge is **into** (or **out of**) a vertex if the edge mark at the vertex is an arrowhead (or a tail). The meaning of standard graph theoretical concepts in directed graphs, such as **parent/child**, **(directed) path**, **ancestor/descendant**, etc., remains the same in mixed graphs. Furthermore, if there is a bi-directed edge between two vertices $X$ and $Y$ ($X \leftrightarrow Y$), then $X$ is called a **spouse** of $Y$. If there is an undirected edge between $X$ and $Y$ ($X - Y$), then $X$ is called a **neighbor** of $Y$.

**Definition 1.** *A mixed graph is* **ancestral** *if*

*(a1) there is no directed cycle;*

*(a2) if $V_1$ is a spouse of $V_2$ (i.e., $V_1 \leftrightarrow V_2$), then $V_1$ is not an ancestor of $V_2$; and*

*(a3) if $V_1$ is a neighbor of $V_2$ (i.e., $V_1 - V_2$), then $V_1$ has no parents or spouses.*

Obviously DAGs and undirected graphs (UGs) – graphs in which all edges are undirected – meet the definition, and hence are special cases of ancestral graphs. The first condition in Definition 1 is just the familiar one for DAGs. Together with the second condition, they define a nice connotation of arrowheads – that is, an arrowhead implies non-ancestorship. The third condition requires that there is no edge into any vertex in the *undirected* component of an ancestral graph. This property simplifies parameterization and fitting of ancestral graphs (RS 2002, Drton and Richardson 2003), and still allows selection effect to be properly represented.

Mixed graphs encode conditional independence relations by essentially the same graphical criterion as the well-known *d-separation* for DAGs, except that in mixed graphs colliders can arise in more edge configurations than they do in DAGs. Given a path $p$ in a mixed graph, a non-endpoint vertex $V$ on $p$ is called a **collider** if the two edges incident to $V$ on $p$ are both into $V$, otherwise $V$ is called a **non-collider**.

**Definition 2 (m-separation).** *In a mixed graph, a path $p$ between vertices $X$ and $Y$ is* **active** *(***m-connecting***) relative to a set of vertices* $\mathbf{Z}$ *($X, Y \notin \mathbf{Z}$) if*

*i. every non-collider on $p$ is not a member of $\mathbf{Z}$;*

*ii. every collider on $p$ has a descendant in $\mathbf{Z}$.*

$X$ *and* $Y$ *are said to be* **m-separated** *by* $\mathbf{Z}$ *if there is no active path between* $X$ *and* $Y$ *relative to* $\mathbf{Z}$.

The following property is true of DAGs and UGs: if two vertices are not adjacent, then there is a set of vertices that m-separates the two. This, however, is not true of ancestral graphs in general. For example, Figure 2(a) is an ancestral graph that fails this condition: $C$ and $D$ are not adjacent, but no subset of $\{A, B\}$ m-separates them.

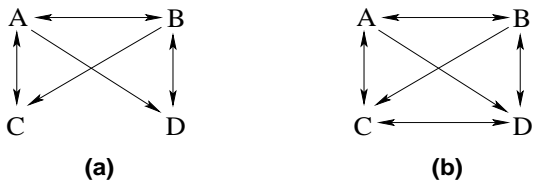

Figure 2: (a) an ancestral graph that is not maximal; (b) a maximal ancestral graph

This motivates the following definition:

**Definition 3 (maximality).** *An ancestral graph is said to be* **maximal** *if for any two non-adjacent vertices, there is a set of vertices that m-separates them.*

It is shown in RS (2002) that every non-maximal ancestral graph can be easily transformed to a *unique* supergraph that is ancestral and maximal by adding bi-directed edges. This justifies considering only those ancestral graphs that are maximal (MAGs). From now on, we focus on maximal ancestral graphs, which we will refer to as MAGs. A notion closely related to maximality is that of inducing path:

**Definition 4 (inducing path).** *In an ancestral graph, a path $p$ between $X$ and $Y$ is called an* **inducing path** *if every non-endpoint vertex on $p$ is a collider and is an ancestor of either $X$ or $Y$.*

For example, in Figure 2(a), the path $\langle C, A, B, D \rangle$ is an inducing path between $C$ and $D$. RS (2002) proved that the presence of an inducing path is necessary and sufficient for two vertices not to be m-separated by any set. So an ancestral graph is maximal if and only if there is no inducing path between any two non-adjacent vertices in the graph.

As shown in RS (2002), the class of MAGs is closed under marginalization and conditioning. Hence MAGs can represent independent relations among observed variables entailed by a DAG with latent confounders (to be marginalized over) and/or selection variables (to be conditioned upon). For details of how MAGs represent DAG models with latent variables, we refer readers to RS (2002) and also Spirtes and Richardson (1996).

## 2.2 MARKOV EQUIVALENCE

As a probabilistic model, a MAG represents a set of joint distributions that satisfy the conditional independence relations implied by m-separation in the MAG. Hence two MAGs that share the same m-separation structures represent the same set of distributions.

**Definition 5 (Markov equivalence).** *Two MAGs $\mathcal{G}_1, \mathcal{G}_2$ (with the same set of vertices) are* **Markov equivalent** *if for any three disjoint sets of vertices $\mathbf{X}, \mathbf{Y}, \mathbf{Z}$, $\mathbf{X}$ and $\mathbf{Y}$ are m-separated by $\mathbf{Z}$ in $\mathcal{G}_1$ if and only if $\mathbf{X}$ and $\mathbf{Y}$ are m-separated by $\mathbf{Z}$ in $\mathcal{G}_2$.*

It is well known that two DAGs are Markov equivalent if and only if they have the same adjacencies and the same unshielded colliders (Verma and Pearl 1990). (A triple $\langle X, Y, Z \rangle$ is said to be **unshielded** if $X, Y$ are adjacent, $Y, Z$ are adjacent but $X, Z$ are not adjacent.) The conditions are still necessary for Markov equivalence between MAGs, but are not sufficient. For two MAGs to be equivalent, some shielded colliders have to be present in both or neither of the graphs. The next definition is related to this.

**Definition 6 (discriminating path).** *In a MAG, a path between $X$ and $Y$, $p = \langle X, \cdots, W, V, Y \rangle$, is a* **discriminating path** *for $V$ if*

  i. *$p$ includes at least three edges (i.e., at least four vertices as specified);*

 ii. *$V$ is adjacent to an endpoint $Y$ on $p$; and*

iii. *$X$ is not adjacent to $Y$, and every vertex between $X$ and $V$ is a collider on $p$ and is a parent of $Y$.*

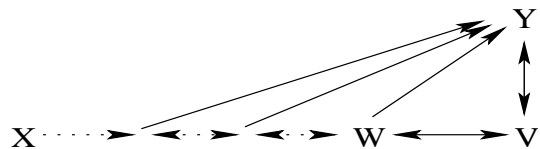

Figure 3: A discriminating path for $V$: the triple $\langle W, V, Y \rangle$ is "discriminated" to be a collider here.

Discriminating paths behave similarly to unshielded triples in that if $p = \langle X, \cdots, W, V, Y \rangle$ is discriminating for $V$, then $\langle W, V, Y \rangle$ is a (shielded) collider (See Figure 3) if and only if every set that m-separates $X$ and $Y$ does not contain $V$; it is a non-collider if and only if every set that m-separates $X$ and $Y$ contains $V$. The following is proved in Spirtes and Richardson (1996).

**Proposition 1.** *Two MAGs over the same set of vertices are Markov equivalent if and only if*

*(e1) They have the same adjacencies;*

(e2) They have the same unshielded colliders;

(e3) If a path $p$ is a discriminating path for a vertex $Y$ in both graphs, then $Y$ is a collider on the path in one graph if and only if it is a collider on the path in the other.

## 2.3 PARTIAL ANCESTRAL GRAPHS

Given a MAG $\mathcal{G}$, we denote its Markov equivalence class, the set of MAGs Markov equivalent to $\mathcal{G}$, by $[\mathcal{G}]$. An edge mark in $\mathcal{G}$ is said to be **invariant** if the mark is the same in all members of $[\mathcal{G}]$.

**Definition 7 (PAG).** *Given a MAG $\mathcal{G}$, the **partial ancestral graph (PAG)** for $[\mathcal{G}]$, $\mathcal{P}_\mathcal{G}$, is a graph with (possibly) three kinds of edge marks: arrowheads, tails, and circles (and hence six kinds of edges: —, →, ↔, ∘—, ∘—∘, ∘→), such that*

  i. *$\mathcal{P}_\mathcal{G}$ has the same adjacencies as $\mathcal{G}$ (and hence any member of $[\mathcal{G}]$) does;*

  ii. *A mark of arrowhead is in $\mathcal{P}_\mathcal{G}$ if and only if it is invariant in $[\mathcal{G}]$; and*

  iii. *A mark of tail is in $\mathcal{P}_\mathcal{G}$ if and only if it is invariant in $[\mathcal{G}]$.*[1]

The mark of circle is obviously intended to represent an edge mark that is not invariant. We will henceforth use *partial mixed graphs* (PMGs) to refer to graphs that may contain circles. Earlier definitions of PAGs in the literature (e.g., Spirtes et al. 1997) do not require PAGs to reveal all invariant marks. Ali et al. (2005) used what is called *joined graphs* to represent $[\mathcal{G}]$. However, a joined graph represents only invariant arrowheads and does not distinguish invariant tails from variant marks.[2] Clearly the most informative representation of a Markov equivalence class of MAGs is the PAG as defined here. The question is how to construct $\mathcal{P}_\mathcal{G}$ from $\mathcal{G}$.

## 3 ARROWHEAD COMPLETE ORIENTATION RULES

To construct $\mathcal{P}_\mathcal{G}$, we start with a graph $\mathcal{P}$ that has the same adjacencies as $\mathcal{G}$ and no informative marks but circles.[3] Then we apply a set of orientation rules that change some circles into arrowheads or tails. In light of $(e2)$ in Proposition 1, we first mark unshielded colliders.

$\mathcal{R}0$ For every unshielded triple of vertices $\langle \alpha, \beta, \gamma \rangle$, if it is an unshielded collider in $\mathcal{G}$, then orient the triple as $\alpha *\!\!\rightarrow \beta \leftarrow\!\!* \gamma$.

($*$ is a meta-symbol that serves as a wildcard for edge marks.) The soundness of $\mathcal{R}0$ is obvious given Proposition 1. That is, after we apply $\mathcal{R}0$ to $\mathcal{P}$, all resulting arrowheads are invariant. In general, however, there are more invariant arrowheads. The following rules are sufficient to identify all:

$\mathcal{R}1$ If $\alpha *\!\!\rightarrow \beta \circ\!\!-\!\!* \gamma$, and $\alpha$ and $\gamma$ are not adjacent, then orient the triple as $\alpha *\!\!\rightarrow \beta \rightarrow \gamma$.

$\mathcal{R}2$ If $\alpha \rightarrow \beta *\!\!\rightarrow \gamma$ or $\alpha *\!\!\rightarrow \beta \rightarrow \gamma$, and $\alpha *\!\!-\!\!\circ \gamma$, then orient $\alpha *\!\!-\!\!\circ \gamma$ as $\alpha *\!\!\rightarrow \gamma$.

$\mathcal{R}3$ If $\alpha *\!\!\rightarrow \beta \leftarrow\!\!* \gamma$, $\alpha *\!\!-\!\!\circ \theta \circ\!\!-\!\!* \gamma$, $\alpha$ and $\gamma$ are not adjacent, and $\theta *\!\!-\!\!\circ \beta$, then orient $\theta *\!\!-\!\!\circ \beta$ as $\theta *\!\!\rightarrow \beta$.

$\mathcal{R}4$ If $p = \langle \theta, ..., \alpha, \beta, \gamma \rangle$ is a discriminating path between $\theta$ and $\gamma$ for $\beta$, and $\beta \circ\!\!-\!\!* \gamma$; then if $\beta \rightarrow \gamma$ appears in $\mathcal{G}$, orient $\beta \circ\!\!-\!\!* \gamma$ as $\beta \rightarrow \gamma$; otherwise orient the triple $\langle \alpha, \beta, \gamma \rangle$ as $\alpha \leftrightarrow \beta \leftrightarrow \gamma$.

$\mathcal{R}1 - \mathcal{R}3$ are essentially Meek's orientation rules in the context of DAGs (Meek 1995). $\mathcal{R}4$ makes use of discriminating paths, and is peculiar to MAGs with bi-directed edges. $\mathcal{R}0 - \mathcal{R}4$ are equivalent to the set of orientation rules given in Ali et al. (2005), except that the latter is formulated in the framework of Joined Graphs which do not distinguish between tails and circles. The results in Ali et al. (2005) entail that $\mathcal{R}0 - \mathcal{R}4$ are sound and complete for constructing invariant arrowheads (also see Zhang 2006, Theorem 3.2.1 and Theorem 3.3.1).

## 4 TAIL COMPLETENESS

Let $\mathcal{P}_{FCI}$[4] be the graph resulting from an exhaustive application of $\mathcal{R}0 - \mathcal{R}4$ to $\mathcal{P}$. In general, $\mathcal{P}_{FCI}$ is not yet the PAG of $[\mathcal{G}]$. In other words, although $\mathcal{P}_{FCI}$ reveals all invariant arrowheads and some invariant tails (due to $\mathcal{R}1$ and $\mathcal{R}4$), some circles therein may hide invariant tails. The goal of the present paper is to supply more tail orientation rules to construct $\mathcal{P}_\mathcal{G}$, the PAG of $[\mathcal{G}]$.

To introduce the extra tail orientation rules, we need a couple of definitions.

---

[1] Zhang (2006) uses the name *complete* or *maximally oriented* PAGs. We will simply call them PAGs in this paper.

[2] Basically we get the joined graph for $[\mathcal{G}]$ by turning all circles in the PAG into tails.

[3] The adjacencies can be constructed even if we are not given a MAG, but instead given a set of independence facts. See Spirtes et al. (1999). Similar comments apply to $\mathcal{R}0$ and $\mathcal{R}4$ below.

[4] We call it $\mathcal{P}_{FCI}$ because $\mathcal{R}0 - \mathcal{R}4$ are essentially the rules used in the FCI algorithm given in Spirtes et al. (1999).

**Definition 8 (uncovered path).** *In a PMG, a path $p = \langle V_0, \cdots, V_n \rangle$ is said to be **uncovered** if for every $1 \leq i \leq n-1$, $V_{i-1}$ and $V_{i+1}$ are not adjacent, i.e., every consecutive triple on the path is unshielded.*

**Definition 9 (potentially directed path).** *In a PMG, a path $p = \langle V_0, \cdots, V_n \rangle$ is said to be **potentially directed** (abbreviated as **p.d.**) from $V_0$ to $V_n$ if for every $0 \leq i \leq n-1$, the edge between $V_i$ and $V_{i+1}$ is not into $V_i$ or out of $V_{i+1}$.*

Intuitively, a p.d. path is one that could be oriented into a directed path by changing the circles on the path into appropriate tails or arrowheads. As we shall see, uncovered p.d. paths play an important role in locating invariant tails. A special case of a p.d. path is where every edge on the path is of the form ∘—∘. We call such a path a **circle path**.

Here is the first block of additional rules:

$\mathcal{R}5$ For every (remaining) $\alpha \circ\!\!-\!\!\circ \beta$, if there is an uncovered circle path $p = \langle \alpha, \gamma, \cdots, \theta, \beta \rangle$ between $\alpha$ and $\beta$ s.t. $\alpha, \theta$ are not adjacent and $\beta, \gamma$ are not adjacent, then orient $\alpha \circ\!\!-\!\!\circ \beta$ and every edge on $p$ as undirected edges (—).

$\mathcal{R}6$ If $\alpha -\!\!\!-\beta \circ\!\!-\!\!*\gamma$ ($\alpha$ and $\gamma$ may or may not be adjacent), then orient $\beta \circ\!\!-\!\!*\gamma$ as $\beta -\!\!\!-\!\!*\gamma$.

$\mathcal{R}7$ If $\alpha -\!\!\circ \beta \circ\!\!-\!\!*\gamma$, and $\alpha, \gamma$ are not adjacent, then orient $\beta \circ\!\!-\!\!*\gamma$ as $\beta -\!\!\!-\!\!*\gamma$.

The pictorial illustrations of $\mathcal{R}5 - \mathcal{R}7$ are given in Figure 4. These rules are related to undirected edges. $\mathcal{R}5$ lead to undirected edges, and $\mathcal{R}6$ depend on undirected edges. So if the given MAG $\mathcal{G}$ does not contain undirected edges, these two are not needed. In that case, moreover, $\mathcal{R}7$ will not get triggered at all, because neither $\mathcal{R}0 - \mathcal{R}4$ introduced earlier nor $\mathcal{R}8 - \mathcal{R}10$ to be introduced shortly can lead to —∘ edges, which are in the antecedent of $\mathcal{R}7$.

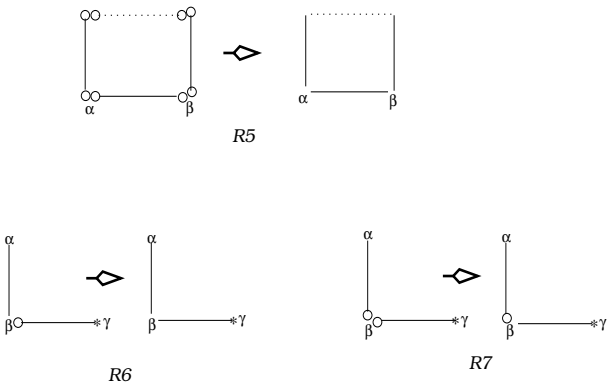

Figure 4: Graphical illustrations of $\mathcal{R}5 - \mathcal{R}7$

That is why we introduce these three rules as a block. Only when there are undirected edges in $\mathcal{G}$ do we need to include $\mathcal{R}5 - \mathcal{R}7$. Recall that undirected edges are motivated by the need to represent selection effects. So if there is no issue of selection bias, we would only consider MAGs with directed and bi-directed edges, in which case $\mathcal{R}5 - \mathcal{R}7$ can be ignored. The next block of rules, by contrast, may still be applicable.

$\mathcal{R}8$ If $\alpha \rightarrow \beta \rightarrow \gamma$ or $\alpha -\!\!\circ \beta \rightarrow \gamma$, and $\alpha \circ\!\!\rightarrow \gamma$, orient $\alpha \circ\!\!\rightarrow \gamma$ as $\alpha \rightarrow \gamma$.

$\mathcal{R}9$ If $\alpha \circ\!\!\rightarrow \gamma$, and $p = \langle \alpha, \beta, \theta, \cdots, \gamma \rangle$ is an uncovered p.d. path from $\alpha$ to $\gamma$ such that $\gamma$ and $\beta$ are not adjacent, then orient $\alpha \circ\!\!\rightarrow \gamma$ as $\alpha \rightarrow \gamma$.

$\mathcal{R}10$ Suppose $\alpha \circ\!\!\rightarrow \gamma$, $\beta \rightarrow \gamma \leftarrow \theta$, $p_1$ is an uncovered p.d. path from $\alpha$ to $\beta$, and $p_2$ is an uncovered p.d. path from $\alpha$ to $\theta$. Let $\mu$ be the vertex adjacent to $\alpha$ on $p_1$ ($\mu$ could be $\beta$), and $\omega$ be the vertex adjacent to $\alpha$ on $p_2$ ($\omega$ could be $\theta$). If $\mu$ and $\omega$ are distinct, and are not adjacent, then orient $\alpha \circ\!\!\rightarrow \gamma$ as $\alpha \rightarrow \gamma$.

These rules are visualized in Figure 5. All of them are about turning partially directed edges ∘→ into directed ones →, which are valuable for the sake of causal inference because ↔ and → represent different causal information (see Richardson and Spirtes 2003).

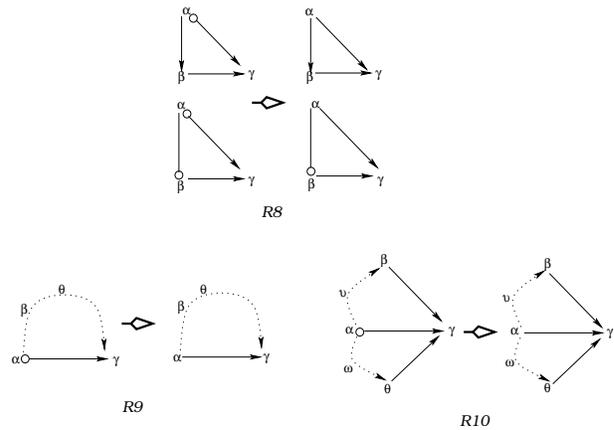

Figure 5: Graphical illustrations of $\mathcal{R}8 - \mathcal{R}10$

There are obviously cases in which the additional rules are applicable. For example, given the MAG in Figure 6(a), $\mathcal{R}0 - \mathcal{R}4$ will give us the graph in 6(b), which miss some invariant tails, but we can apply $\mathcal{R}9$ to get them, as shown in 6(c). In fact, it is not hard to construct cases to show that all the extra rules given above except possibly $\mathcal{R}8$ are necessary. (We do not yet know if $\mathcal{R}8$ is really needed.)

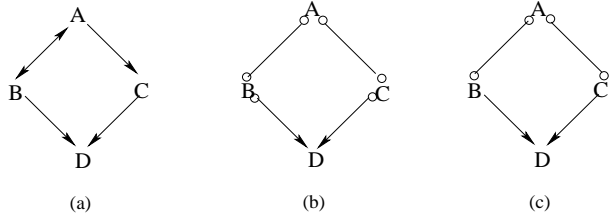

Figure 6: An example where $\mathcal{R}9$ is needed

So in general $\mathcal{R}0 - \mathcal{R}4$ are not able to pick out all invariant tails. Our main result is that $\mathcal{R}5 - \mathcal{R}10$ are sufficient for that purpose. Let $\mathcal{P}_{AFCI}$ (AFCI denotes "Augmented FCI") be the graph resulting from an exhaustive application of $\mathcal{R}5-\mathcal{R}10$ to $\mathcal{P}_{FCI}$. Here is the main theorem:

**Main Theorem** $\mathcal{P}_{AFCI} = \mathcal{P}_{\mathcal{G}}$. In other words, $\mathcal{R}0 - \mathcal{R}10$ are sound and complete for identifying invariant edge marks in $\mathcal{G}$.

Unfortunately, the current proof of the theorem is way too long to present here. We can only explain the main steps of the argument, and refer interested readers to the full proof in Zhang (2006, chapter 4).

First, the soundness of $\mathcal{R}5 - \mathcal{R}10$ is not hard to show.

**Lemma 1.** $\mathcal{R}5 - \mathcal{R}10$ *are sound.*

*Proof.* For each rule, we just need to show that any mixed graph that violates the rule does not belong to $[\mathcal{G}]$.

$\mathcal{R}5$: Note that the antecedent of this rule implies that $\langle \alpha, \gamma, \cdots, \theta, \beta, \alpha \rangle$ forms an uncovered cycle that consists of $\circ\!\!-\!\!\circ$ edges. Suppose a mixed graph, contrary to what the rule requires, has an arrowhead on this cycle. In light of $\mathcal{R}1$, the cycle must be oriented as a directed cycle to avoid unshielded colliders not in $\mathcal{G}$. But then the graph is not ancestral.

$\mathcal{R}6$: if any graph, contrary to what the rule requires, contains $\alpha - \beta \leftarrow\!\!*\gamma$, the graph is not ancestral.

$\mathcal{R}7$: Suppose a mixed graph, contrary to what the rule requires, has an arrowhead at $\beta$ on the edge between $\beta$ and $\gamma$. Then either $\alpha - \beta \leftarrow\!\!*\gamma$ is present, in which case the graph is not ancestral; or $\alpha \to \beta \leftarrow\!\!*\gamma$ is present, in which case the graph contains an unshielded collider not in $\mathcal{G}$.

$\mathcal{R}8$: This rule is analogous to $\mathcal{R}2$. Obviously if a mixed graph, contrary to what the rule requires, contains $\alpha \leftrightarrow \gamma$, then either an almost directed cycle is present or there is an arrowhead into an undirected edge, and hence the graph is not ancestral.

$\mathcal{R}9$: The essentially same argument for the soundness of $\mathcal{R}5$ applies here.

$\mathcal{R}10$: The antecedent of the rule implies that the triple $\langle \mu, \alpha, \omega \rangle$ is not a collider in $\mathcal{G}$, which means at least one of the two edges involved in the triple is out of $\alpha$ in any MAG in $[\mathcal{G}]$. Suppose a graph in $[\mathcal{G}]$, contrary to what the rule requires, contains $\alpha \leftrightarrow \gamma$. Then the edge(s) out of $\alpha$ must be a directed edge for the graph to be ancestral. It follows that either $p_1$ or $p_2$ is a directed path in the graph to avoid unshielded colliders not in $\mathcal{G}_T$. In either case, $\alpha$ is an ancestor of $\gamma$, and hence the graph is not ancestral, a contradiction. $\square$

Completeness is much harder. We need to show that for every circle in $\mathcal{P}_{AFCI}$, there is a MAG in $[\mathcal{G}]$ in which the corresponding mark is an arrowhead, and there is one in $[\mathcal{G}]$ in which the corresponding mark is a tail. The latter is of course taken care of by the arrowhead completeness result. To show the former, the following lemma is needed.

**Lemma 2.** *The following properties hold of $\mathcal{P}_{AFCI}$:*

**P1** *for any three vertices $A, B, C$, if $A*\!\!\to B \circ\!\!-\!\!*C$, then there is an edge between $A$ and $C$ with an arrowhead at $C$, namely, $A*\!\!\to C$. Furthermore, if the edge between $A$ and $B$ is $A \to B$, then the edge between $A$ and $C$ is either $A \to C$ or $A\circ\!\!\to C$ (i.e., it is not $A \leftrightarrow C$).*

**P2** *For any two vertices $A, B$, if $A -\!\!\circ B$, then there is no edge into $A$ or $B$.*

**P3** *if $A -\!\!\circ B \circ\!\!-\!\!\circ C$, then $A -\!\!\circ C$; if $A -\!\!\circ B \circ\!\!\to C$, then $A \to C$ or $A \circ\!\!\to C$.*

**P4** *There is no such cycle as $A -\!\!\circ B -\!\!\circ C -\!\!\circ \cdots -\!\!\circ A$.*

*Proof Sketch:* **P1** is analogous to Lemma 1 of Meek (1995). It is related to $\mathcal{R}0-\mathcal{R}4$, and is the key property needed to prove arrowhead completeness. See Lemma 4.1 of Ali et al. (2005) and Lemma 3.3.1 of Zhang (2006). **P2**−**P4** are related to $\mathcal{R}5-\mathcal{R}7$. The basic idea is to show that $\mathcal{R}5 - \mathcal{R}7$ cannot lead to configurations violating **P2** − **P4**. See proofs of Lemmas 4.3.1, 4.3.3 and 4.3.4 in Zhang (2006). $\square$

Let the *circle component* of $\mathcal{P}_{AFCI}$, denoted by $\mathcal{P}^c_{AFCI}$, be the subgraph of $\mathcal{P}_{AFCI}$ consisting of all $\circ\!\!-\!\!\circ$ edges in $\mathcal{P}_{AFCI}$.

**Lemma 3.** *For any edge $A \circ\!\!-\!\!\circ B$ in $\mathcal{P}^c_{AFCI}$, $\mathcal{P}^c_{AFCI}$ can be oriented into a DAG with no unshielded colliders in which $A \to B$ appears.*

*Proof Sketch:* Given Lemma 5 in Meek (1995), it suffices to show that $\mathcal{P}^c_{AFCI}$ is chordal. And it is easy to show, given properties **P1** and **P3**, that if $\mathcal{P}^c_{AFCI}$ is non-chordal, then there is a chordless cycle consisting of ∘—∘ edges in $\mathcal{P}_{AFCI}$, which should have been oriented by $\mathcal{R}5$. □

Properties **P1** – **P4** are very useful in proving the following important fact:

**Lemma 4.** *Let H be the graph resulting from the following procedure applied to $\mathcal{P}_{AFCI}$:*

(1) *Orient the circles on ∘→ edges in $\mathcal{P}_{AFCI}$ as tails, and orient the circles on —∘ edges in $\mathcal{P}_{AFCI}$ as arrowheads (that is, turn all ∘→ edges and all —∘ edges into directed edges →); and*

(2) *orient $\mathcal{P}_{AFCI}$ into a DAG with no unshielded colliders.*

*Then H is a MAG and is Markov equivalent to G.*

*Proof Sketch:* **P1** ensures that turning ∘→ into → will not create configurations in violation of conditions (a1)-(a2) in Definition 1 or an inducing path in violation of maximality. **P2** and **P3** ensure that after turning —∘ into →, **P1** still holds. **P4** ensures that no directed cycle would result from turning —∘ into →. **P1** then ensures that no matter how one orient a ∘—∘ edge, no directed cycle or unshielded collider that involves already existing arrowheads would be created. Given all these, it is relatively easy to check that $H$ satisfies the definition of MAG, and satisfies the conditions in Proposition 1 with $\mathcal{G}$. □

Lemma 4 has a couple of important implications. First, together with Lemma 3, it implies that for any circle on a —∘ edge or a ∘—∘ edge in $\mathcal{P}_{AFCI}$, there is a member in $[\mathcal{G}]$ in which the corresponding mark is an arrowhead. Hence, no circle on —∘ or ∘—∘ edges in $\mathcal{P}_{AFCI}$ corresponds to an invariant tail. Second, it is clear that no extra undirected edges or bi-directed edges are introduced in constructing $H$ in Lemma 4. This means that in any Markov equivalence class of MAGs, there is a representative in which all undirected edges and bi-directed edges are invariant. This fact was used in Zhang and Spirtes (2005) to establish a transformational characterization of Markov equivalence between directed MAGs. Moreover, a MAG with the fewest number of undirected edges and bi-directed edges is probably easier to fit, in light of the fact that UGs are harder to fit than DAGs and the results presented in Drton and Richardson (2004). Lemma 4 gives a simple way of constructing such a representative from a PAG. By contrast, the algorithm given in Ali et al. (2005) does not construct a representative with the fewest undirected edges.

What is left to show is that every circle on a ∘→ edge also corresponds to an arrowhead in some member of $[\mathcal{G}]$. (In other words, $\mathcal{R}8 - \mathcal{R}10$ have picked up all invariant tails that were hidden in ∘→ edges.) This turns out to be the most difficult step. The space here only permits us to give a very rough skeleton of the proof. The basic strategy consists of two major steps. Let $J \circ\!\!\rightarrow K$ be an arbitrary ∘→ edge in $\mathcal{P}_{AFCI}$. In the first step, we show that we can orient $\mathcal{P}^c_{AFCI}$ — the circle component of $\mathcal{P}_{AFCI}$ — into a DAG with no unshielded colliders that satisfies certain conditions relative to $J \circ\!\!\rightarrow K$. This DAG orientation of $\mathcal{P}^c_{AFCI}$ together with operation (1) in Lemma 4 yield a MAG Markov equivalent to $\mathcal{G}$.

In the second step, we make use of a result on equivalence-preserving mark changes presented in Zhang and Spirtes (2005) and Tian (2005), and argue that the particular MAG constructed in the first step can be transformed into a MAG containing $J \leftrightarrow K$ through a sequence of equivalence-preserving changes of → into ↔. (The conditions put down in the first step for the DAG orientation of $\mathcal{P}^c_{AFCI}$ play a crucial role in proving this.) It then follows that the resulting MAG with $J \leftrightarrow K$ is also Markov equivalent to $\mathcal{G}$, which gives us what we need.

The first step does most of the work, and also occupies most of the proof. In particular, the constraints (defined relative to an edge $J \circ\!\!\rightarrow K$) on the DAG orientation of $\mathcal{P}^c_{AFCI}$ are quite complicated, and it takes a lot of effort to show they can indeed by satisfied. Again, interested readers have to consult Zhang (2006) for the details, which we hope to simplify soon. Here we will just state the result.

**Lemma 5.** *For every $J \circ\!\!\rightarrow K$ in $\mathcal{P}_{AFCI}$, there is a MAG in $[\mathcal{G}]$ in which the edge appears as $J \leftrightarrow K$.*

This follows immediately from Corollary 4.3.33 in Zhang (2006). Our main theorem follows from Lemmas 1, 4, 5 and the arrowhead completeness result.

## 5 CONCLUSION

We have provided a characterization of Markov equivalence classes for MAGs — which are well suited to represent DAGs with latent confounders and/or selection variables — in the style of Meek (1995)'s characterization for DAGs. The characterization is by way of a set of orientation rules that are sound and complete for constructing commonalities among MAGs in a Markov equivalence class. We also showed how to construct a representative MAG with fewest number of undirected edges and bi-directed edges from a PAG, which is potentially useful for scoring PAGs in score-based PAG search.

Our results are directly relevant to the constraint-based approach to causal discovery in the presence of latent confounders and selection variables. As mentioned in footnote 4, the FCI algorithm (Spirtes et al. 1999), which takes an oracle of conditional independence facts as input, essentially makes use of $\mathcal{R}0 - \mathcal{R}4$ (replacing the references to the given MAG with references to the oracle) in the orientation stage. Whether FCI is complete has been an open problem for a while. We now know that if the oracle is reliable, the FCI algorithm, as it stands, outputs a partial mixed graph that reveals all invariant arrowheads but not all invariant tails in the true causal MAG. Augmented by the additional rules given in this paper, the FCI algorithm becomes complete.

This completeness result is significant because under the causal interpretation of MAGs (Richardson and Spirtes 2003), tails can represent important qualitative causal information. Roughly, $A \rightarrow B$ in a MAG means that $A$ is a cause of $B$ or a cause of a hidden selection variable. This becomes particularly informative when selection effect is known to be absent, in which case $A \rightarrow B$ means that $A$ has a causal influence on $B$.[5] By contrast, if it is only known that there is an arrowhead at $B$ but not known whether the mark at $A$ is a tail, all one can say is that $B$ is not a cause of $A$.

The orientation rules fall naturally into three blocks. $\mathcal{R}0 - \mathcal{R}4$ are arrowhead complete. So if one only cares about invariant arrowheads, the other rules can be ignored. $\mathcal{R}5 - \mathcal{R}7$ are related only to undirected edges, or the presence of selection bias. They are not needed if, as is often the case, one worries about latent confounding but not selection bias. Finally, $\mathcal{R}8 - \mathcal{R}10$ may give us more directed causal arrows.

**Acknowledgement**

I thank Peter Spirtes and Thomas Richardson for checking the proofs in Zhang (2006).

---

[5] Though it is compatible with there also being a latent common cause of $A$ and $B$. Thanks to an anonymous referee for emphasizing this point.